# An Easy yet Effective Method for Detecting Spatial Domain LSB Steganography

**Minati Mishra**
Department of Information and Communication Technology
Fakir Mohan University, Balasore, Odisha, India

**Flt. Lt. Dr. M. C. Adhikary**
Department of Applied Physics and Ballistics
Fakir Mohan University, Balasore, Odisha, India

**ABSTRACT**

Digitization of image was a revolutionary step for the fields of photography and Image processing as this made the editing of images much effortless and easier. Image editing was not an issue until it was limited to corrective editing procedures used to enhance the quality of an image such as, contrast stretching, noise filtering, sharpening etc. But, it became a headache for many fields when image editing became manipulative. Digital images have become an easier source of tampering and forgery during last few decades. Today users and editing specialists, equipped with easily available image editing software, manipulate digital images with varied goals. Photo journalists often tamper photographs to give dramatic effect to their stories. Scientists and researchers use this trick to get theirs works published. Patients' diagnoses are misrepresented by manipulating medical imageries. Lawyers and Politicians use tampered images to direct the opinion of people or court to their favor. Terrorists, anti-social groups use manipulated Stego images for secret communication. In this paper we present an effective method for detecting spatial domain Steganography.

## 1. INTRODUCTION

Significant advancements in digital imaging during the last decade have added a few innovative dimensions to the field of image processing. Steganography and watermarking are few such creative dimensions of image processing those have gained wide popularity among the researchers. Digital image watermarking techniques are generally used for authentication purposes and are achieved by embedding a small piece of information into copyrighted digital information. Steganography, on the other hand, hides a large amount of data secretly into a digital medium and is generally used for secret communications. It is one of the effective means of data hiding that protects data from unauthorized or unwanted disclosure and can be used in various field such as medicines, research, defence and intelligence for secret data storage, confidential communication, protection of data from alteration and disclosure and access control in digital distribution.

Like every coin has two sides, all technological developments are associated with both bad as well as good applications and Steganography is not an exception to this [1]. Though there are many good reasons to use data hiding techniques and these should be used for legitimate applications only but, unfortunately, Steganography can also be used for illegitimate reasons [2]. For instance, someone trying to steal data can conceal it in another file and send it out through an innocent looking email. The information stolen and passed may be a patient's confidential test reports, the tender information of a company/ organization or even the defence plans of a country. No doubt, terrorists and criminals can use this method to secretly spread their action plans and though no evidence is yet established to this claim still, it is claimed that, Steganography was used to pass the execution plan of the 9/11 WTC attack[3]. Since Steganography mainly targets innocent looking digital images for their huge size and popularity therefore it has become an important requirement of the time to differentiate between real innocent images from those innocent looking Stego images. In this paper we are suggesting an easier method that can be used to detect spatial domain LSB Steganography by analyzing the histograms of digital images.

The organization of the paper is as follows. The following section explains the process of Steganography. Steganalysis is discussed in section 3. Experimental results are given in section 4 followed by the summary and conclusion at the end.

## 2. STEGANOGRAPHY

Steganography is a process of secret communication where a piece of information (a secret message) is hidden into another piece of innocent looking information, called a cover. The message is hidden inside the cover in such a way that the very existence of the secret information remains concealed without raising any suspicion in the minds of the viewers [4, 5].

Steganographic procedures make use of several media such as text, audio, video or image as covers but, digital image Steganography is more popular among the researchers as images are more common forms of mediums that are used worldwide for data transmission and also due to their data hiding capacity. The embedding (M) and extraction (X) processes in digital image Steganography are mapping given by the following equations.

$$M : \{ C \times (K) \times S \} \rightarrow C' \quad \text{----------- (eq. 1)}$$

And

$$X : \{C' \times (K)\} \rightarrow S \quad \text{-------- (eq. 2)}$$

Where, M and X are the embedding and extraction functions respectively, C is the cover image, C' is the Stego image, K is an optional set of keys (K in M and X may be same or may be different depending upon the encryption algorithm used), S is a set of secret messages [6].

Just like the image tampering techniques, Steganography is also an image manipulation technique. This too manipulates digital images from their original captures like the image tampering methods but with a different purpose. Tampering techniques manipulate images to fake a fact and mislead the viewer to misbelieve the truth behind a scene whereas Steganography manipulates it for covert communication. Because of its inherent purpose of data hiding, Steganography requires the original and the Stego image to look alike but a tampered images need to look as natural as possible keeping the tampering undetectable to human vision.

Steganography is broadly classified into two categories such as spatial domain Steganography and transform domain Steganography. Spatial domain methods take advantage of the human visual system and directly embed data by manipulating the pixel intensities whereas in the transform domain procedures, the image is first transformed into frequency domain and then the message is embedded. The transform domain procedures are more robust against statistical attacks and manipulations in comparison to the spatial domain methods but spatial domain techniques are more popular due to their simplicity and ease of use.

### 2.1 Cover Image Selection

As imperceptibility and message security are the important criteria of steganography, choice of cover image plays an important role. Images with large areas of uniform color, with a few colors or with a unique semantic content, such as fonts, computer generated arts, charts etc are poor candidates for cover as they will easily reveal the secret content. Although computer-generated fractal images can be considered to be good covers due their complexity and irregularity but as they are generated by strict deterministic rules that may be easily violated by message embedding. Therefore, scans of photographs or images obtained with a digital camera those contain a high number of colors are usually considered safe for Steganography. In general, grayscale images are considered as the best covers by many experts [7]. Again, the raw- uncompressed BMP format images generally are preferred as candidates for covers over the lossy compressed format images such as JPEG, Wavelet, JPEG2000 and palette formats (such as, GIF) as they offer the highest capacity and best overall security [8].

## 2.2 Bit-plane slicing

Digital images can be monochrome (bi-tone), grayscale (true-tone) or color depending upon the permissible intensity levels of each pixel i.e. whether each pixel is represented by only one bit, 8-bits or 24-bits. Like the gray-level ranges, the contribution made by a single bit to the total image appearance is also important for specific applications such as Steganography and watermarking. If each pixel of an image is represented by m bits then the image can be imagined to be composed of m numbers of 1-bit planes ranging from bit-plane 0 through bit-plane m-1. For example, a monochrome image can have only one bit-plane whereas there are 8 bit-planes in a gray scale image and 24 bit-planes (8 bits each, with respect to the 3 channels R, G and B) in a color image. Figure 1 shows the bit-plane representation of a 8x7 pixel gray scale image.

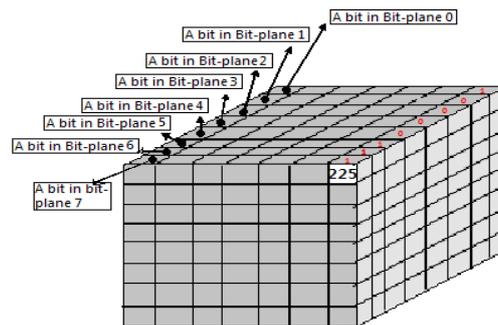

**Figure 1: Bit-plane representation of a gray scale image**

The $0^{th}$ bit-plane or the least significant bit-plane (LSB plane) is the plane that consists of bits with minimum positional value ($2^0$= 1) and the MSB plane (most significant bit plane or the $7^{th}$ bit-plane) consists of all high order bits. Therefore a pixel with gray value 225 can have bits 1,1,1,0,0,0,0,1 in $7^{th}$ to $0^{th}$ bit-planes respectively. Figure 2 shows different bit-planes of the grayscale Lena image.

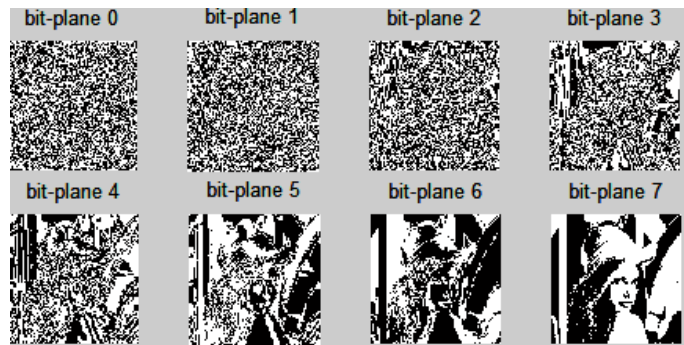

**Figure 2: The 8 bit-planes of Lena image**

Information hiding in Steganography is generally carried out in two steps. Firstly, identifying redundant bits in a cover medium which can be modified without degrading the quality of the cover medium and secondly, selecting a subset of the redundant bits to be replaced with data from secret messages. The stego medium is created by replacing the selected redundant bits with message bits. It can be noticed from the above figure that the higher order bits (especially the top four bits) contain majority of the visually significant image information. The other bits contribute to the finer details in the image. In spatial Steganography, a digital image is generally sliced into different bit-planes and the lower order bit-planes are replaced by secret messages. As the lower-order bits do not carry much visually significant image information, replacement of those bits does not make any visible disruption to the image and the covert message remains secret inside the Stego image. Figure.3 shows the Stego Lena image and its bit planes. It can be seen that the Stego image completely conceals the secret messages inside without revealing anything to the viewers. The Peak signal-to-noise ratio (PSNR) that is considered to be an approximation to human perception quality also remains high, 43.3342 in this case, without raising any suspicion in the minds of the viewers.

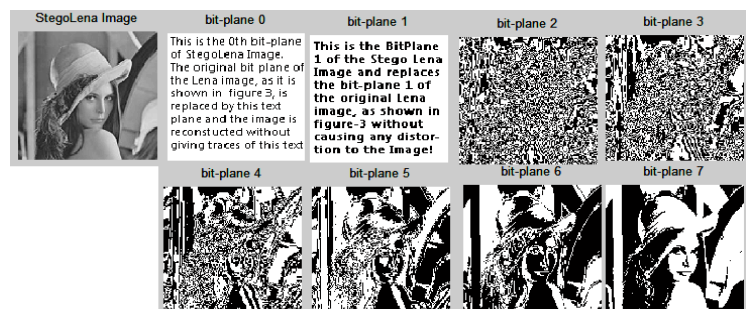

**Figure 3: The Stego Lena image and its 8 bit-planes**

As Stego images look exactly same as the original images therefore, it becomes a challenge for the viewers to differentiate the Stegoes from their original counter parts.

## 3. STEGANALYSIS

Steganalysis is the art and science of detecting messages hidden using Steganography [9]. It can be considered to be a two-class pattern classification problem which aims to determine whether a medium under test is a cover or Stego [2]. Hence, the goal of steganalysis is to identify suspected packages, determine whether or not they have a payload encoded into them, and, if possible, recover that payload. Steganalysis is similar to cryptanalysis with a slight difference. In cryptanalysis, it is obvious that intercepted data contains a message therefore; the task here is just to find the underlying message by decrypting the encrypted text. On the other hand steganalysis generally starts with a huge set of suspected data files without any prior information about whether they contain any hidden message or not and if any secret message is there then which file contains it. Therefore, steganalysis can be viewed to be a two step process. One is to reduce this large set to a smaller subset of files those are most likely to have been altered and the second is to separate the covert signal from the carrier. Detection of suspected files is straightforward when the original, unmodified carriers are available for comparison but when only a single image is available; it becomes a tough problem to say whether it is manipulated or not as steganography attempts to make the Stego indistinguishable from the cover.

Distinguishing a Stego file from that of a real innocent file is a major task in steganalysis. Because of its global nature, Steganography generally leaves detectable traces in the medium's characteristics and careful examination of the modified media can reveal the existence of some secret content in it defeating the very purpose of Steganography, even if the secret content is not exposed. This paper illustrates how Stego images can be easily separated from the real innocent ones through histogram analysis (Histanalysis).

### 3.1 Histogram

A digital image is a two dimensional function f(x, y) where, x and y are spatial coordinates, f is the amplitude at (x, y), also called the intensity or gray level of the image at that point and x, y, f are finite- discrete quantities.

The histogram of a digital image with gray levels from 0 to L-1 is a discrete function $h(r_k)=n_k$, where, $r_k$ is the $k^{th}$ gray level, $n_k$ is the number of pixels in the image with that gray level, n is the total number of pixels in the image and k = 0, 1, 2, …, L-1. [10] In other words, a histogram plot gives the number of counts of each gray level. In a histogram plot, the horizontal axis corresponds to gray level values, $r_k$ and the vertical axis corresponds to the values of $h(r_k)=n_k$ or $p(r_k)=n_k/n$ if the values are normalized. Histograms are the basis to various spatial domain processing and provide useful image statistics. Information inherent in histograms is quite useful for a number of image processing applications. Figure.4 shows a grayscale image and its corresponding histogram.

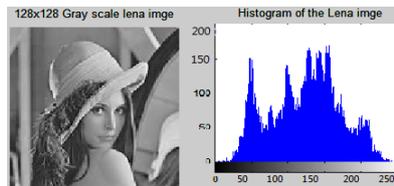

**Figure 4: Grayscale Lena image and the corresponding Histogram**

### 3.2 JPEG Compression

JPEG (Joint Photographic Experts Group) is an international compression standard for continuous-tone grayscale and color still images. JPEG standard has two basic compression methods. The popular DCT-based encoding and decoding method offers lossy compression whereas the predictive method is specified for lossless compression. The ISO/ITU-T standard defines several modes of JPEG compression such as baseline sequential, progressive and hierarchical. In baseline mode, the image is divided into 8x8 non-overlapping blocks, each of which is discrete cosine transformed. The transformed block coefficients are quantized with a uniform scalar quantizer, zig-zag scanned and entropy coded with Huffman coding. Whereas, in case of interlaced progressive JPEG format, data is compressed in multiple passes of progressively higher detail. This is ideal for large images that will be displayed while downloading over a slow connection, allowing a reasonable preview after receiving only a portion of the data. Figure 5 shows the block diagrams of baseline sequential JPEG encoding and decoding steps. [11]

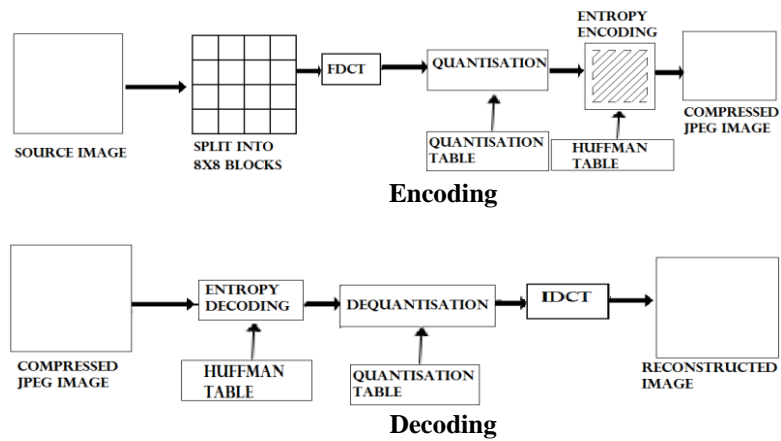

**Encoding**

**Decoding**

**Figure 5: Baseline sequential JPEG encoding and decoding process**

### 3.2.1 Discrete Cosine Transform

Given an image *f(x, y)* of size *NxN*, the forward discrete transform *F (u, v)* and given the *F (u, v)*, the inverse discrete transform *f(x, y)* can be obtained by the general relations:

$$F(u,v) = \sum_{x=0}^{N-1} \sum_{y=0}^{N-1} f(x,y) g(x,y,u,v)$$

$$f(x,y) = \sum_{x=0}^{N-1} \sum_{y=0}^{N-1} F(u,v) h(x,y,u,v)$$

Where, *g(x, y, u, v)* and *h(x, y, u, v)* are called the forward and inverse transform kernels respectively. In case of discrete cosine transformation (DCT), both the forward and inverse transform kernels are given by a single relation [8]

$$g(x,y,u,v) = h(x,y,u,v) = \alpha(u)\alpha(v) \cos\left[\frac{(2x+1)u\pi}{2N}\right] \cos\left[\frac{(2y+1)v\pi}{2N}\right]$$

Where,

$$\alpha(k) = \begin{cases} \sqrt{\frac{1}{N}} & \text{if } k = 0 \\ \sqrt{\frac{2}{N}} & \text{if } k = 1,2.. \ N-1 \end{cases}$$

In a DCT transformed block *F (u, v)*, the upper left corner bit *F (0,0)* represents the DC component and rest of the *F (u, v)* are called AC components. Because significant signal energy of an image lies in the low frequency DC components, those appear in the upper left corner of the DCT blocks, and since the lower right values representing higher frequencies are small enough and are often neglected without causing much visible distortion to the image therefore, DCT offers compression.

### 3.2.2 Quantization

It is the process of approximating a continuous signal by a set of discrete signals and in this step corresponding quantized blocks *C (i, j)* of the DCT coefficient blocks *D (i, j)* are obtained using the formula

$$C(i,j) = round\left(\frac{D(i,j)}{Q(i,j)}\right)$$

Where Q (i, j) is a quantization table of selected quality. After this quantization process, values of high frequency AC coefficients of the DCT block are usually become zero providing a lossy compression. The non-zero DC coefficients of each block are then coded separately using Hoffman's loss-less entropy encoding algorithm. Given in the Figure 6 (b) are the DCT coefficients of the matrix A (given in Figure 6(a) after quantized by $Q_{50}$ (Figure. 6(c) and then rounded up.

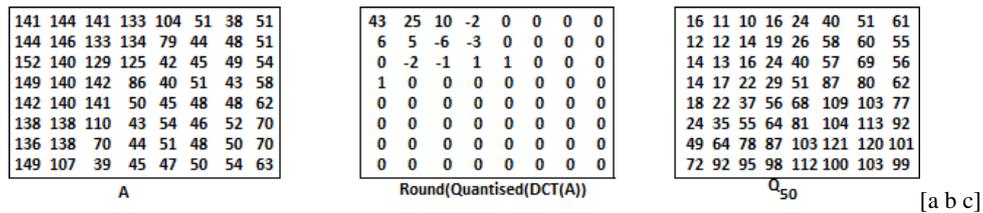
[a b c]

**Figure 6: A 8 x 8 matrix and it's coefficients after DCT and Quantization steps**

## 4. EXPERIMENTAL RESULTS AND DISCUSSIONS

Natural photographs being true tone images, the gray levels in these images generally show a continuous variation between a minimum and maximum gray value. For example, in the above Lena image of figure 4, we have pixels with every gray value between 11 and 242. But when one or more bit-planes of an image are replaced by some other data, the gray counts for certain bins increase/ decrease in random giving rise to a type of discontinuity to the gray value counts. In the following Figure.7, image (a) is an original image (without secret message embedded) captured in PNG format and figure (b) shows the histogram of gray level counts of this image where x-axis represents the bins and y-axis the gray level counts. Figure (e) shows the histogram of the same image after saving it in JPEG. In both these histograms (as given in (b) and (e) in the figure below), it can be seen that the gray counts vary continuously between the minimum to maximum bin. Figure.7 (c) is a stego image which is formed by replacing the LSB bits of the image (a) with secret message and saved in TIFF format.

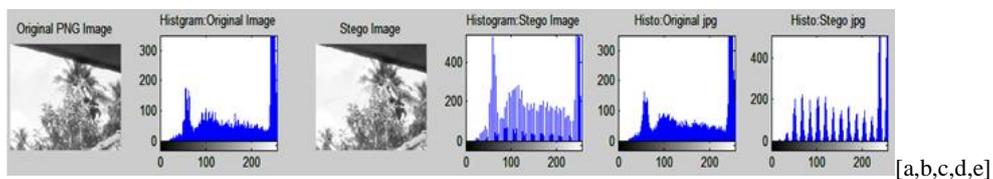
[a,b,c,d,e]

**Figure 7: Histograms of original and Stego images saved in PNG, TIFF and JPEG**

The histogram plot of this image is given in Figure.7 (d) and it is clear from the plot that the gray counts for certain bins have increased (e.g., for some bins between 50 and 100) whereas for some other bins that have decreased introducing a type discontinuity to the histogram plot that seems as if two different histograms are super imposed with each other. This discontinuity is further increased as shown in Figure.7 (e), when the Stego image is saved in

International Journal of Computer Science and Business Informatics

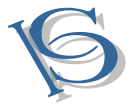

IJCSBI.ORG

JPEG format and then the histogram is plotted. The gray level counts have been clearly partitioned into different gray zones and the histogram shows a clear sign of discreetness with alternate picks and valleys which is not an usual phenomena for photographic images.

To establish this finding, we plotted the histograms of a number of gray scale images by replacing one, two and three lower order bit-planes of those with text messages and then saving the stego image both in .tiff and .jpg formats. Since the luminance component of a color image is equivalent to a grayscale image and since grayscale images are generally considered suitable for steganography, we have used grayscale images for our experiments. MATLAB have been used for bit-slicing, embedding and creating the .TIF stego images. The TIFF images are then converted to .jpg using Windows paintbrush tool. Histograms are plotted using the MATLAB's *imhist* function giving the bins in x-axis and gray counts in y-axis. Figure.8 shows the results of this operation performed on the 128x128 grayscale Lena image.

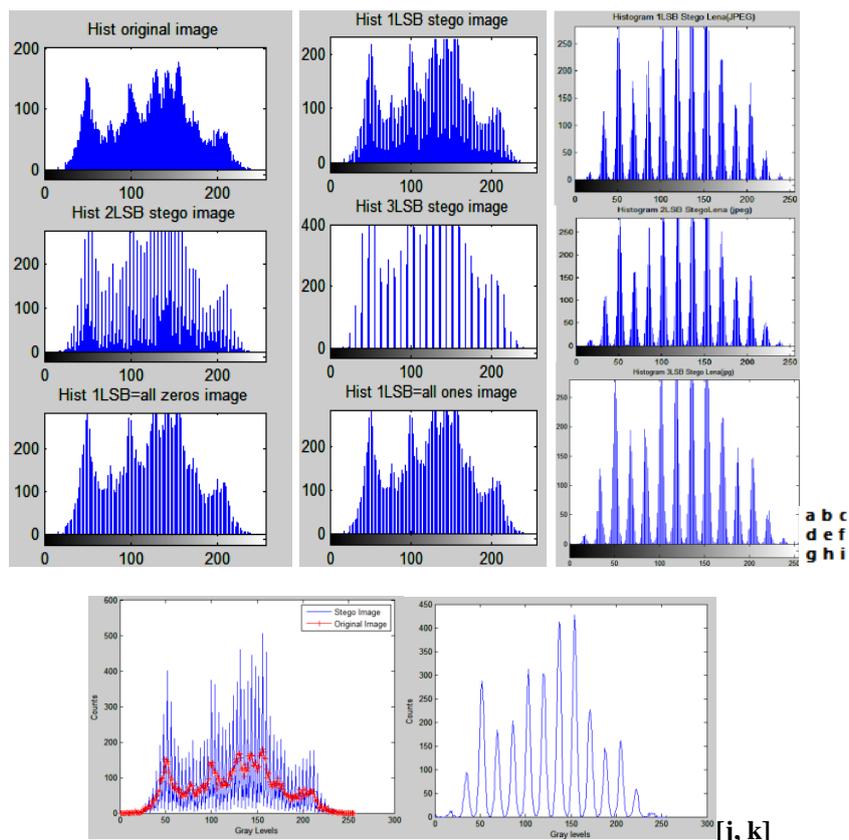

[j, k]

**Figure 8: Histograms of 128x128 bit Lena (tiff) image without Stego and with Stego performed.**

**International Journal of Computer Science and Business Informatics**

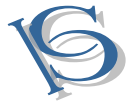

IJCSBI.ORG

Figure.8 (a) shows the histogram of the original Lena (tiff) image, b is the histogram of stegolena after the LSB plane is replaced with text and saved in .tif format, c shows the histogram of LSB replaced stegolena saved in JPEG format. d, e, f and h show the histograms of 2 bit-plane and 3-bit plane replaced stegolena images (TIFF) and their JPEG counterparts respectively. Histogram counts of the original Lena image and that of the Stegolena are plotted in j and those of the stegolena after saving it in JPEG is given in Figure.8 (k).

Though JPEG images with high compression ratio (low quality factor) also produce histograms of the type Figure.8 (b) but as it is discussed in section 2.1, JPEG images are considered to be very poor candidates for spatial Steganography and are generally not considered for this purpose. Secondly, after data is embedded into bit planes, Stego images are never saved in lossy compression formats such as JPEG as these formats discard the lowers order bits (so also the secret messages embedded into those bits!) in order to achieve compression. Thirdly, uncompressed natural images never show a histogram pattern as that of Figure. 8(b) or 8(d). Therefore, when a raw format image such as BMP, TIFF produces a histogram with such an unexpected pattern, it can be suspected to be a Stego medium.

## 5. SUMMARY AND CONCLUSION

Today, digital images not only provide forged information but also work as agents of secret communication. With the availability of a wide range of easy Steganographic methods, these popular digital media are used for secret data transmission, sometimes with legitimate goals and sometimes for immoral purposes. Lots of work has been done on steganalysis and tamper detection techniques and still researchers worldwide are working to successfully detect manipulations made to digital images. In this paper we have used an extremely easy but highly effective histanalysis method to detect spatial domain digital image Steganography.